\documentclass[traditabstract]{aa} 
\usepackage{graphicx}
\usepackage{color}
\usepackage{natbib}
\usepackage{txfonts}
\usepackage{enumitem}

\def\fgrst{\textit{Fermi Gamma-Ray Space Telescope}}
\def\fermilat{\textit{Fermi}/LAT}

\def\gama{$\gamma$-ray }

\begin{document}
\title{The connection between gamma-ray emission and millimeter flares
  in \fermilat\ blazars}

    \author{J. Le\'on-Tavares\inst{1}
            \and 
            E. Valtaoja\inst{2}
            \and
            M. Tornikoski \inst{1}
            \and
             A. L\"ahteenm\"aki\inst{1}
            \and E. Nieppola \inst{1,3}}

    \institute{ Aalto University Mets\"ahovi Radio Observatory,  Mets\"ahovintie 114, FIN-02540
    Kylm\"al\"a, Finland. \\ 
    \email{leon@kurp.hut.fi} \and
   Tuorla Observatory, Department  of Physics and Astronomy, University of Turku, 20100 Turku, Finland. \and
   Finnish Center for Astronomy with ESO (FINCA), University of Turku, V\"ais\"al\"antie 20, FI-Piikki\"o, Finland.}

 \abstract{ We compare the $\gamma$-ray photon flux  variability of northern blazars in the \fermilat\ First Source Catalog with 37 GHz radio flux density curves from the Mets\"ahovi quasar  monitoring program. We find that the relationship between simultaneous  millimeter (mm) flux density and $\gamma$-ray  photon flux is different for different types of blazars. The flux  relation between the two bands is positively correlated for quasars and does no exist for BLLacs. Furthermore, we find that the levels of $\gamma$-ray emission in high states depend on the phase of the high frequency radio flare, with the brightest $\gamma$-ray events coinciding with the initial stages of a mm flare. The mean observed delay from the beginning  of a mm flare to the  peak of the $\gamma$-ray emission  is about 70 days, which places the average location of the $\gamma$-ray production at or downstream of the radio core. We discuss  alternative scenarios for the production of $\gamma$-rays at distances of parsecs along the length of the jet.  }

 
   \titlerunning{The connection between $\gamma$-rays  and mm flares.}

   \maketitle

%

\section{Introduction}

A relativistic jet is a clear taxonomical  characteristic of  extragalactic sources  detected in $\gamma$-rays. Sources with jets pointing close to our line of sight  are called  blazars and are the brightest and  most dominant population of active galactic nuclei (AGN)  in the $\gamma$-ray sky \citep[e.g.][]{Fichtel_1994,1fgl}.  The radiation mechanism  of the $\gamma$-ray emission in blazars is widely believed  to be inverse Compton scattering of ambient  photons, either from inside the jet (synchrotron-self-Compton or SSC; e.g. Bloom \& Marscher 1986),  or from a source external to the jet (external Compton scattering or EC), where the source of seed photons could be  the accretion disk  \citep[e.g.][]{dermer_1993}, the broad-line region  \citep[e.g.][]{sikora_1994}, or  perhaps the dusty  torus \citep[e.g.][]{blazejowski_2000}. We refer to  \citet{boettcher_review_2010} for a  review of theoretical models for blazar emission.  Despite all the
theoretical modeling efforts, the  precise location in general of the $\gamma$-ray emission within sources  is still controversial, which in turn  makes  the origin of  the seed photons for inverse-Compton scattering unclear. The proposed models can be roughly divided into two categories: those with $\gamma$-rays originating relatively close to the black hole and the accretion disk, inside the broad-line region (BLR), and those with $\gamma$-rays originating in the radio jet, at distances of several parsecs and well beyond the BLR.

Studies based on data from the \emph{EGRET} instrument onboard the Compton Gamma Ray Observatory triggered an open debate  about the location of the \gama emission site in blazars. The most popular opinion was that $\gamma$-rays are produced within the BLR region via EC (e.g. Sikora et al. 1994). However, other studies found that high levels of \gama  emission occurred  after the ejections of superluminal jet components  \citep{jorstad_2001},  and that $\gamma$-ray detected sources tend to have ongoing high frequency radio flares \citep{valtaoja_1995,valtaoja_1996,anne_2003}. These results led the authors  to conclude that strong $\gamma$-ray emission in blazars  was produced  in growing shocks in the relativistic jets at  parsec-scale  distances  from the black hole.    Well beyond the central BLR,   the only source of seed photons appears to be   the jet itself,  implying that SSC is the main radiation mechanism  for the  strongest  \gama flares in blazars.  For a historical perspective of the results obtained during the \emph{EGRET} era, we refer to   \citet{aller_2010}

The dramatically improved $\gamma$-ray data from the Large Area Telescope (LAT) onboard the \fgrst\
has opened up the possibility  of testing results obtained from the \emph{EGRET} era regarding the origin of $\gamma$-rays. Several studies, based
on the first year of  LAT  operations,   have shown that:  (i)   the $\gamma$-ray  and the averaged radio flux densities  are significantly correlated \citep{kovalev_2009,giroletti_2010,ghirlanda_2010,mahony_2010,nieppola_2010,angelakis_2010,richards_2010,linford_2011,arshakian_2011}, and (ii) blazars detected at $\gamma$-rays are more likely to have larger Doppler factors \citep{lister_2009,savolainen_2010,tornikoski_2010} and larger apparent opening angles \citep{pushkarev_2009} than those not detected by LAT.  This observational evidence strongly suggests  that  radio and $\gamma$-ray  emission  have a co-spatial origin.

To locate and identify the region where  the bulk of $\gamma$-ray emission is produced, and to provide details about  its connection to  the radio jet,  an analysis of  simultaneous radio and $\gamma$-ray light curves is necessary. \citet{pushkarev_2010},  using   data from the  MOJAVE  survey  \citep{lister_mojave} and the monthly binned $\gamma$-ray light curves provided by the 11-month LAT catalogue \citep{1fgl}, reported  that radio flux-density variations lag significantly behind those in $\gamma$-rays, with  delays ranging from one to eight months (in the observer's frame). This suggests that there is a  correlation between the parsec-scale radio emission and the strength of the $\gamma$-rays.  However, the method employed by the authors  did not allow them to clearly characterize the sequence and the structure of individual flares.

Furthermore, we highlight two caveats about the interpretation of radio/gamma correlation analyses, which have often been interpreted too simplistically. First, it is  well-known  that there is usually a considerable delay between mm and cm radio flares. Thus, although cm-flares would tend to peak after the $\gamma$-ray flares, mm-flares would show shorter delays or possibly even peak before the $\gamma$-rays. The very important second caveat is that a correlation analysis tends to measure the distance between the peaks, especially if the flares have different timescales (as the radio and the $\gamma$-ray flares tend to have). However, a radio flare starts to grow a considerable time before it peaks. The \emph{beginning} of a millimeter flare  coincides with the ejection of a new VLBI component from the radio core  (e.g., Savolainen et al. 2002). This is the epoch that must be compared with the $\gamma$-ray flaring, not the epoch of the radio flare maximum. The crucial question is whether a $\gamma$-ray flare occurs \emph{before} the beginning of a mm-flare, or \emph{after} it; in the former case the $\gamma$-rays originate upstream of the radio core (the beginning of the radio jet), in the latter, they originate downstream of the radio core, presumably from the same disturbance (i.e. shock) that is visible in the radio regime.  (As we discuss in Section 4, the finite length of both the radio core and the disturbance passing through it makes the real situation slightly more complicated. However, the size of the radio core is small compared to its distance from the black hole / accretion disk, and also  compared to the distance the newly created disturbance travels during the rise and the peaking of the mm-flare.)


\begin{table}
  \caption[]{The sample of 60 northern radio sources used in this work. Only 45 sources had   an adequate decomposition of the total flux density curves at  37~GHz. }
\begin{tabular}{lllllllcccccc}
 \hline \hline
 
  Source    &  Alias  &Type& Phase   &  t$_{0}^{mm}$-t$_{peak}^{LAT}$ & Distance \\
                    &           &          &               &  [days]                                                & [$ pc $] \\
       (1)       &   (2)     & (3)    & (4)        &   (5)                                                     &(6)  \\             
 \hline\hline
 0048-097        &                        & BLO  &   0.8  & -58.90  &        ...   \\
 0059+581       &                        & QSO &   1.1  & -79.32  &      6.44 \\
 0106+013       &                        & HPQ &   0.4  & -63.00  &      8.94 \\
 0109+224       &S2\ 0109+22 &  BLO &   0.6  & -28.54  &          ... \\
 0133+476       &                        & HPQ &   1.1  &  -62.15  &     8.36 \\
 0212+735       &                        & HPQ &   1.1  &   -88.33 &     1.44 \\
 0218+357       &                        & QSO &   0.9  &   -74.68  &          ... \\
 0219+428       &3C\ 66            & BLO  &   0.9  &   -73.08  &   ... \\
 0234+285       &                        & HPQ &  ...      & ...             &   ...\\
 0235+164       &                        & BLO  &   0.6  &   -29.03  &     3.60 \\
 0316+413       &3C\ 84            & GAL  &   0.6  &   -37.50  &     0.03 \\
 0336-019        &CTA\ 026       & HPQ &   0.8  &   -55.96  &    10.18 \\
 0420-014        &                        & HPQ &   0.5  &   -33.08  &     3.22 \\
 0440-003        &NRAO\ 190   & HPQ &   0.3  &    16.36  &   ... \\
 0507+179       &                        & QSO &  1.1  &   -76.11  &   ... \\
 0528+134       &                        & HPQ &  0.6  &   -34.29  &     6.45 \\
 0716+714       &                        & BLO  &  ...     &    ...         &   ...           \\
 0736+017       &                        & HPQ &  1.4  &  -104.29  &    10.50 \\
 0754+100       &                        & BLO &  0.6  &   -44.45  &     3.54 \\
 0805-077        &                        &QSO  &  ...     &    ...         &   ...          \\
 0814+425       &                        &BLO  &  ...     &    ...         &   ...          \\
 0827+243       &OJ\ 248          & LPQ &  1.6  &  -143.26  &    20.05 \\
 0829+046       &                        & BLO  &  ...     &    ...         &   ...         \\
 0851+202       &OJ\ 287          & BLO &  0.1  &    68.44  &    11.60 \\
 0917+449       &                        & QSO&  0.5  &   -30.59  &   ... \\
J0948+0022   &                         &QSO &   ...     &    ...         &   ...          \\
 1055+018       &                         &HPQ &      0.7  &   -63.45  &     3.79 \\
 1101+384       &MRK\ 421       & BLO &  ...     &    ...         &   ...          \\
 1118-056        &                         &QSO&  ...     &    ...         &   ...          \\
 1156+295       & 4C\ 29.45       & HPQ &     1.2  &   -81.55  &    28.22 \\
 1219+285       & ON\ 231          & BLO  &     1.5  &  -132.64  &   ... \\
 1222+216       &PKS1222+21  & LPQ  &     1.2  &  -125.24  &    17.33 \\
 1226+023       &3C\ 273            & LPQ  &     1.3  &  -207.59  &    35.13 \\
 1253-055        &3C\ 279            & HPQ &     0.8  &   -50.11  &    13.46 \\
 1308+326       &                           & BLO &     0.7  &  -101.08  &    20.69 \\
 1324+224       &                           &QSO&  ...     &    ...         &   ...          \\
 1406-076        &PKS\ 1406-07  & QSO &    1.1  &   -74.82  &   ... \\
 1502+106       &PKS\ 1502+10 & HPQ&     1.5  &  -106.83  &     5.70 \\
 1510-089        &PKS\ 1510-08  & HPQ &    0.4  &    -0.98  &     0.46 \\
 J1522+3144  &                            &QSO&  ...     &    ...         &   ...        \\
 1606+106       &                           & LPQ  &     1.1  &   -84.31  &    15.67 \\
 1633+382       &4C\ 38.41          & HPQ &     1.6  &  -151.59  &    30.56 \\
 1641+399       &3C\ 345             & HPQ &     0.4  &   -34.60  &     3.95 \\
 1652+398       &MRK\ 501          & BLO &  ...     &    ...         &   ...        \\
 J1700+6830   &                            &  QSO&  ...     &    ...         &   ...       \\
 1717+177       &PKS\ 1717+17  &HPQ &      0.4  &   -23.63  &   ... \\
 1739+522       &S4\ 1739+52     &HPQ &      0.7  &   -65.27  &   ... \\
 1749+096       &PKS\ 1749+096 &BLO&      1.7  &  -153.02  &    10.03 \\
 1803+784       & S5\ 1803+784  & BLO &      ...     &    ...         &   ...       \\                               
 1823+568       &4C\ 56.27            &BLO &      1.2  &  -317.90  &    38.39 \\
 J1849+6705  &                              & QSO&  ...     &    ...         &   ...       \\
 2022-077        &PKS\ 2022-077  &QSO &      1.0  &   -74.10  &   ... \\
 2141+175     &                              &LPQ&  ...     &    ...         &   ...       \\
 2144+092       &                             &QSO &      1.8  &  -483.94  &   ... \\
 2200+420       &BL\ Lac               &BLO &      0.9  &   -45.12  &     2.95 \\
 2201+171       &                             &QSO &      1.7  &  -170.63  &   ... \\
 2227-088        &                             &HPQ &      1.1  &   -66.17  &     5.06 \\
 2230+114       &                             &HPQ &      1.4  &  -116.37  &    11.46 \\
 2234+282       &                             &HPQ &      0.9  &  -101.39  &   ... \\
 2251+158       &3C\ 454.3           & HPQ&     0.7  &   -28.98  &     8.19 \\
\hline
\end{tabular}
 \tablefoot{-- Columns are as follows: (1) source name; (2) alias;  (3) spectroscopic type,  (4) the phase of the mm flare at the time of the $\gamma$-ray peak (0 - 1 growing phase,  1- 2 decaying phase); (5) the observed time delay between the onset of a mm flare and the $\gamma$-ray peak; (6)   the linear distance between the region  of   mm-flare inception  (i.e. radio-core) and the location of the $\gamma$-ray emission region.}
\end{table}



  \begin{figure*}
   \includegraphics[width=\textwidth]{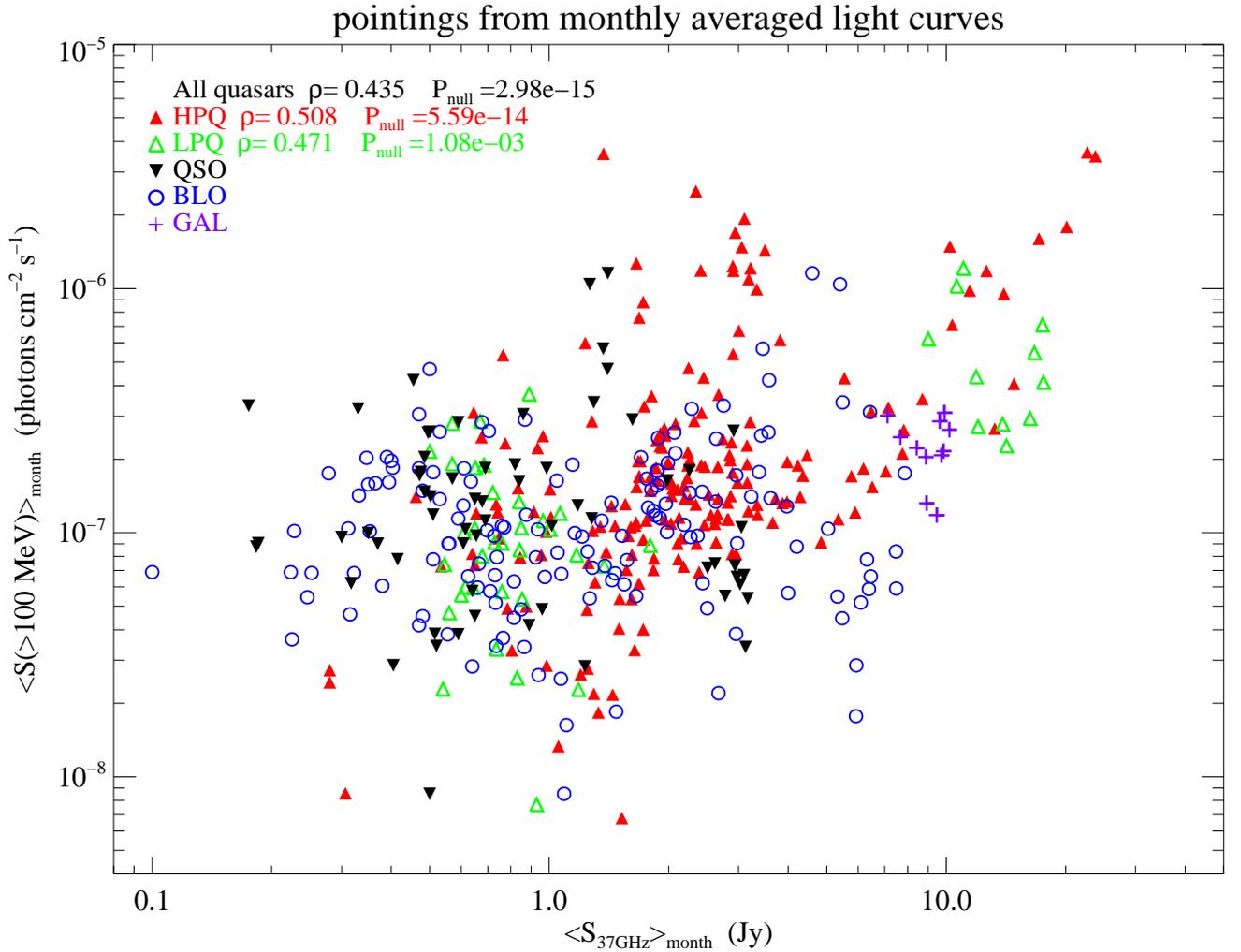}
   \caption{The simultaneous flux - flux relation for the combined sample of
     60 radio-loud AGN analyzed here. The different types of sources
     are symbol coded as shown in the legend. The correlation
     coefficients are shown only when the significance is $\geq 99.9~\%$}
    \end{figure*}

In this paper, we combine  the finely  sampled  37~GHz Mets\"ahovi  light curves and the monthly binned $\gamma$-ray light curves provided by  the \fermilat\ First Source Catalog \citep[][1FGL]{1fgl}. By using a radio flare decomposition method, we determine the beginning epochs of millimeter flares  (cf. equation 3 in Section 4) and their phases during $\gamma$-ray flaring events  to establish the true temporal sequence between $\gamma$-ray and radio flaring. In the following, we use a $\Lambda$CDM cosmology with values within $1\sigma$ of the WMAP results \citep{komatsu_2009}; in particular,   H$_{0}$=71 km s$^{-1}$ Mpc$^{-1}$, $\Omega_{m}=0.27$ and  $\Omega_{\Lambda}=0.73$.

\section{Comparison of Mets\"ahovi and \fermilat\  light curves}

The Mets\"ahovi quasar monitoring program \citep{terasranta_1998} currently includes about 250 AGN at 37~GHz.  From them, we selected a sample of sources that fulfill the following criteria: (i)  well-sampled light curves during the period 2007-2010, covering the 1FGL period, (ii)  a firm association with the 1FGL catalogue, and (iii) a $\gamma$-ray monthly light curve during the 1FGL period that is significantly different from a flat one.

Our final sample consists of 60 sources  classified according to their optical spectral type as highly polarized quasars (HPQ, 22), low polarization quasars (LPQ,~5), quasars without any  optical polarization data (QSO,~15), BL~Lac type objects (BLO,~17), and radio galaxies (GAL,~1). When comparing the averaged radio and $\gamma$-ray properties of the 1FGL sources, and the dependence of the $\gamma$-ray detection likelihood on radio properties,  we refer to  \citet{tornikoski_2010} and \citet{nieppola_2010}. The sample of sources studied in this work is listed in Table 1 along with their spectral types.

To compare radio flux densities  and $\gamma$-ray photon fluxes, monthly binned radio light curves were created from the Mets\"ahovi flux density curves at 37~GHz. The time bins are the same as in the 1FGL flux  history curves, allowing us to compare simultaneously $\gamma$-ray photon flux  and radio flux-density variations with the time resolution of a month. To categorize the phase of the mm flares at the time of the $\gamma$-ray maxima, we  decomposed the total flux density  variations at 37~GHz into a small number of exponential flares as in previous studies  \citep[e.g][]{valtaoja_1999,savolainen_2002,anne_2003,hovatta_2009}.

\section{Results}

For each of the 60 sources, we were able to investigate the relation between the $\gamma$-ray photon flux  and 37~GHz flux densities during the 11-month 1FGL period. However, only for a subsample of 45 sources  did we achieve an adequate decomposition of the total flux density curves, allowing us to  categorize in detail the radio flare state during the $\gamma$-ray maxima.

    
\begin{figure*}
   \includegraphics[width=\textwidth]{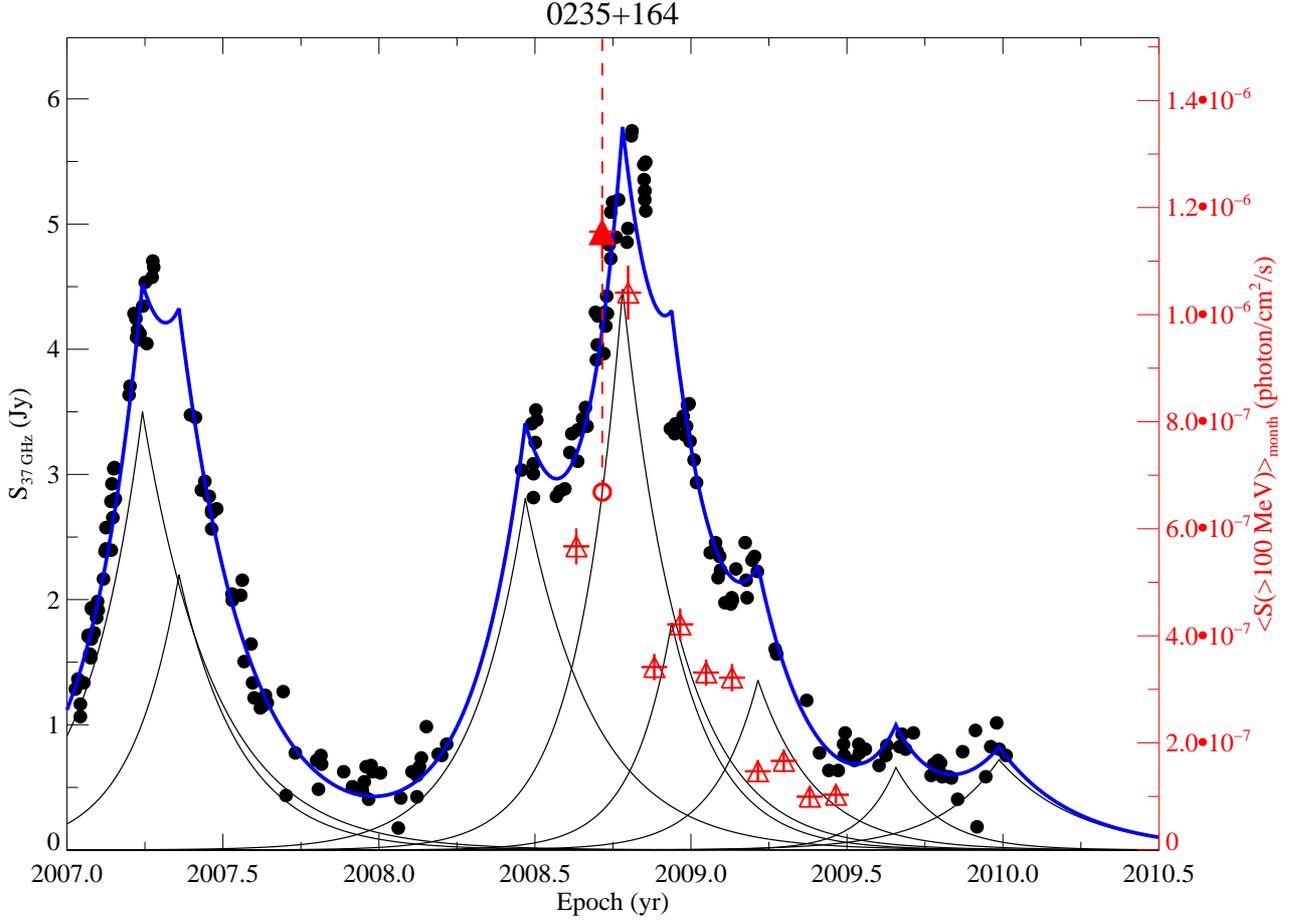}
   \caption{The mm and 1FGL light curves for 0235+164, shown by
     filled circles and triangles, respectively. The total flux density
     curve at 37~GHz has been decomposed using individual
     exponential flares following the method described in
     \citet{valtaoja_1999}.  The filled triangle represents the most
     significant peak in the $\gamma$-ray flux density during the 1FGL
     period. The dashed vertical line is drawn to highlight the
     relation between the peak in the $\gamma$-ray light curve and
     the ongoing millimeter flare.  }
    \end{figure*}

  \begin{figure*}
   \includegraphics[width=\textwidth]{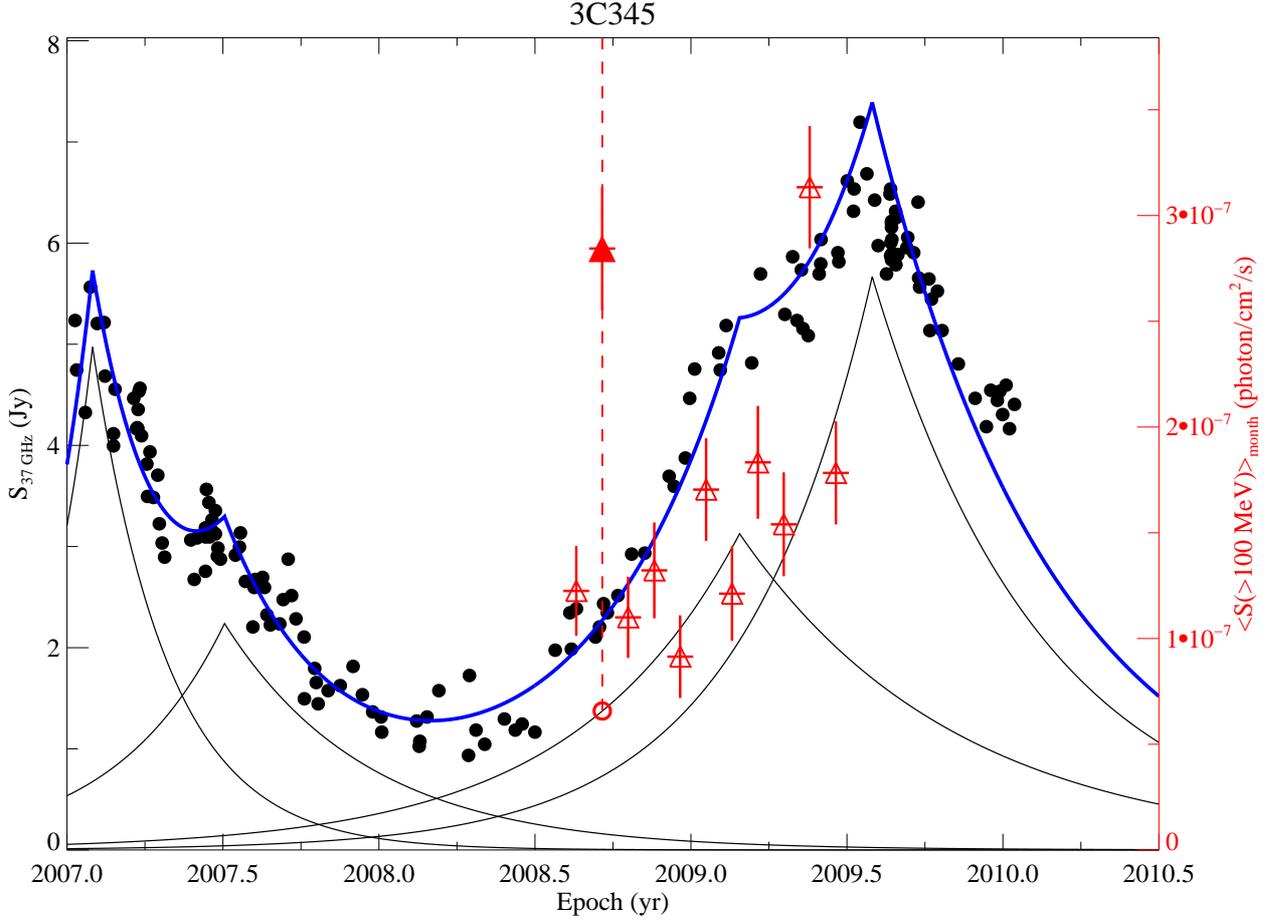}
   \caption{The recent flux  history at 37~GHz (filled circles) and  $\gamma$-rays (triangles) of 3C~345.  The
     filled triangle represents the most significant peak in the   $\gamma$-ray flux density during the 1FGL period, which is clearly associated with a growing shock and a mm flare.  }
    \end{figure*}

\begin{figure*}
   \includegraphics[width=\textwidth]{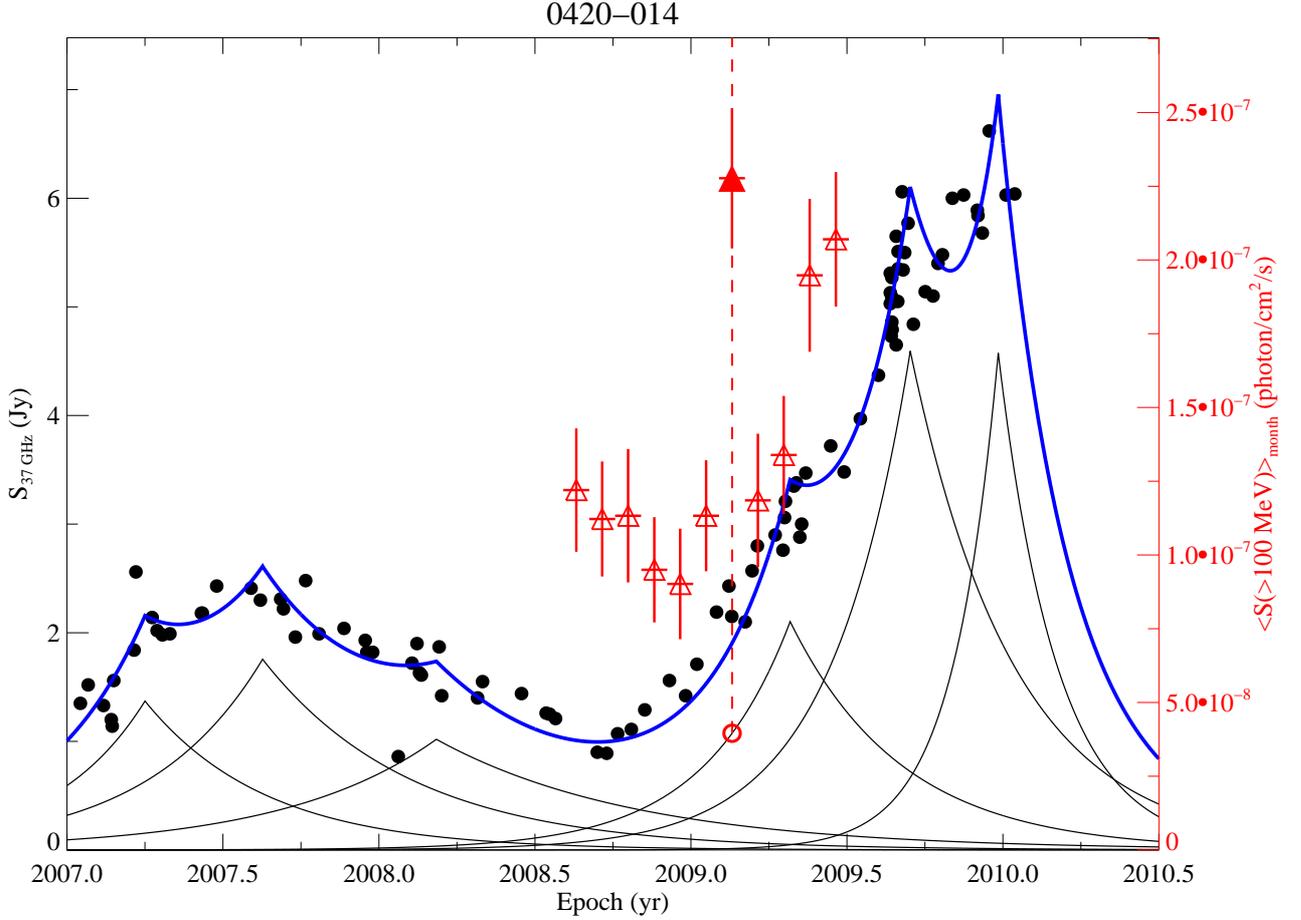}
   \caption{The millimeter and the $\gamma$-ray flux  evolution of the highly polarized quasar  0420-014 (PKS 0420-01). A strong  millimeter flare is evident after almost  two years  of  quasi-quiescent  activity at mm wavelengths, accompanied by a rise in $\gamma$-rays. Although there is a  general trend between the mm and gamma ray flux densities, the highest level of $\gamma$-ray emission  was reached when the first mm flare was  rising (flare phase = 0.5, see Table 1).    }
    \end{figure*}

    \begin{figure*}
   \includegraphics[width=\textwidth]{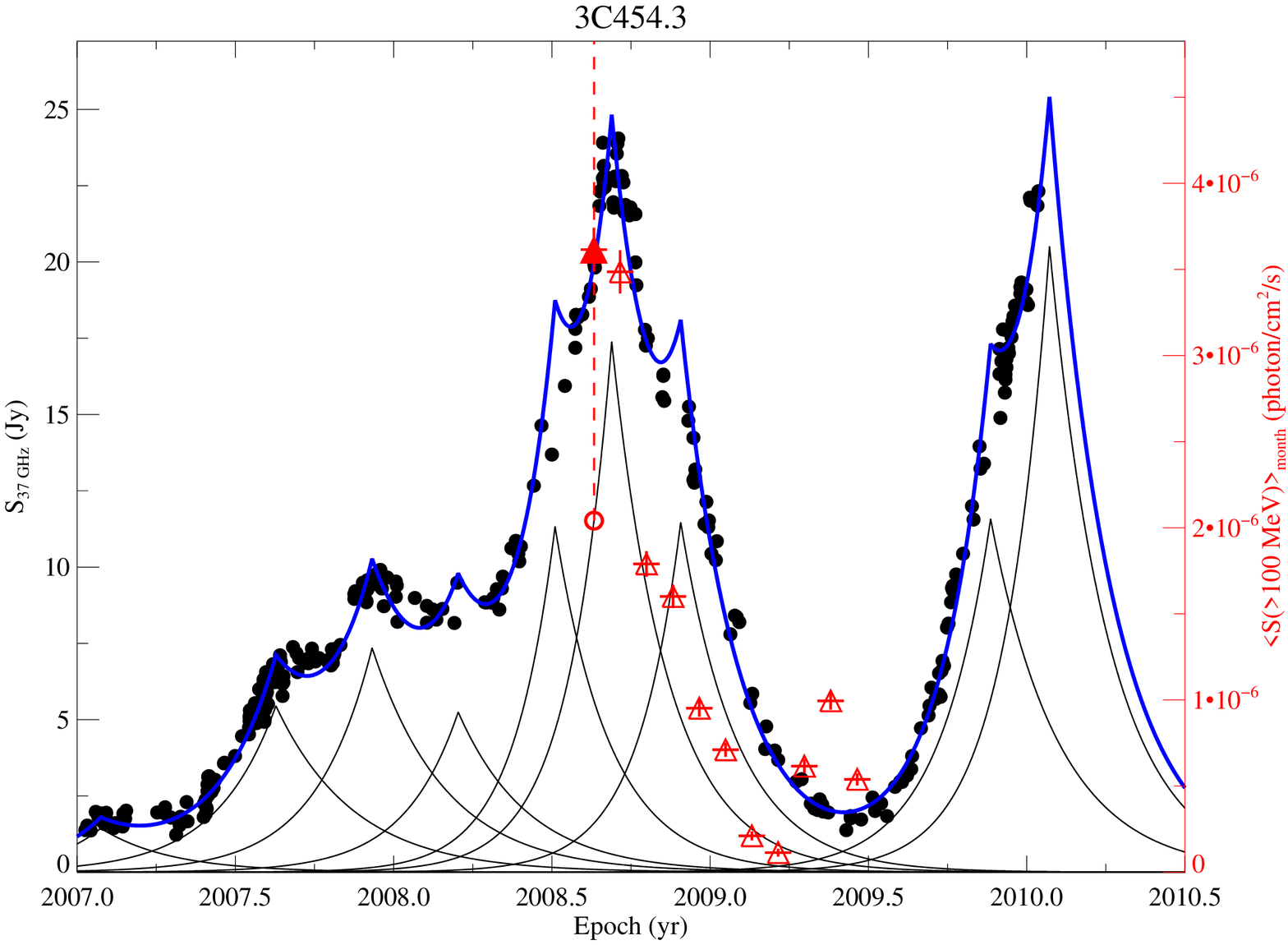}
   \caption{ The monthly binned $\gamma$-ray light curve for 3C~345.3. The strongest $\gamma$-ray flux density is reached during the rise of an mm radio flare. Note the weaker $\gamma$-ray flare around 2009.4, discussed in the text. A $\gamma$-ray outburst has been reported in December 2009  \citep{ackermann_3c4543_2010}, outside the 1FGL period, occurring close to the peak of the latest strong radio flare. }
    \end{figure*}
    

\subsection{Simultaneous photon flux - flux density relation}

Using the  finely sampled light curves from the Mets\"ahovi quasar monitoring program, we compare simultaneous $\gamma$-ray  fluxes  and 37~GHz flux densities. Figure 1 shows that simultaneous measurements at the two bands appear to be positively correlated. However, we find significant differences between quasars and BL~Lacs, which we describe below.  By applying the Spearman's rank correlation test, two very clear results emerge. First, there is a significant positive correlation between the $\gamma$-ray  photon flux and the 37~GHz flux density for quasars, while the BL~Lac fluxes  are not correlated. Second, the strength and the significance of the correlations is different for each type of quasar.

The photon flux - flux density  correlation for quasars is absent for QSOs, significant ($\rho= 0.47, P_{null}= 99.9\%$) for LPQs, and very significant ($\rho= 0.50, P_{null} > 99.9\%$) for HPQs. Such a dependence on the degree of optical polarization may arise naturally  if the polarization indicates the viewing angle of the jet, with sources with high optical polarization having their jets oriented closest to our line of sight \citep[e.g.][]{hovatta_2009}. The dependence of the flux - flux relation on optical polarization agrees with previous results, where it has been shown that the brightest $\gamma$-ray emitters have preferentially smaller viewing angles \citep{valtaoja_1995,anne_2003}  and consequently higher Doppler boosting factors \citep{lister_2009,savolainen_2010,tornikoski_2010}. Since $\gamma$-ray fluxes and radio flux densities are significantly correlated for sources where the relativistic jet is aligned close to our line of sight, this implies that there is  a strong coupling between the radio and the $\gamma$-ray emission mechanisms. That the correlation is seen on monthly timescales further indicates a cospatial origin in quasar-type blazars.

 \begin{figure}
    \includegraphics[width=\columnwidth]{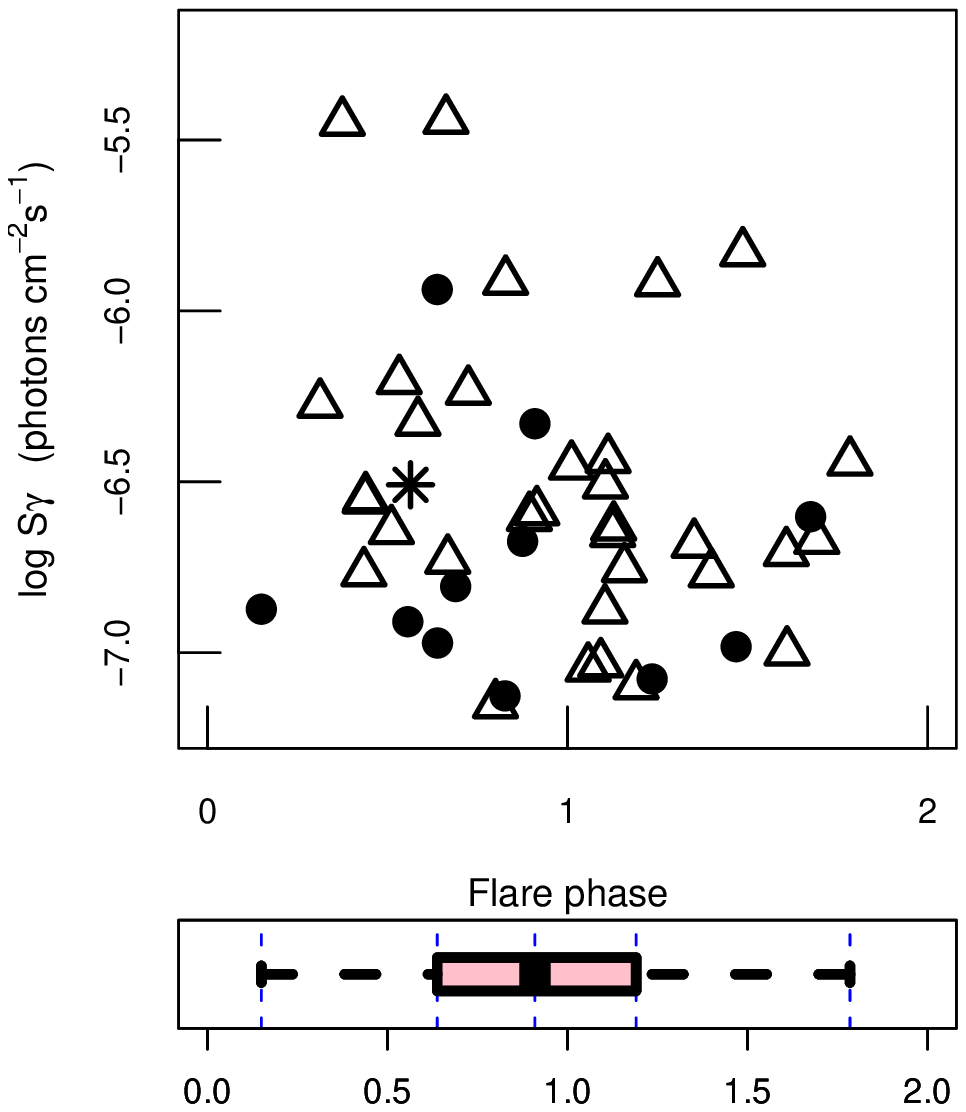}
    \caption{The maximum gamma ray flux density versus the radio flare phase (0,
      beginning; 1 peak; 2 end of the flare). The bottom panel shows
      the distribution of the flare phases in a boxplot, with the
      median located at early stages of the flare. Quasars, BLLacs,  and
      radio-galaxies are shown by triangles, circles, and asterisks,
      respectively.}
      \label{fig:phase}
     \end{figure}

\subsection{Connections between ongoing flares and high states of $\gamma$-ray emission}

We  decomposed the 37~GHz light curves into individual exponential flares, each of which corresponds to a new disturbance  created in the jet that is often detectable as a new VLBI component (Valtaoja et al. 1999, Savolainen et al. 2002). We compared the phase of the individual flares with the $\gamma$-ray light curves obtained from the 1FGL catalogue  to determine whether high states of $\gamma$-ray emission are associated with disturbances propagating downstream of the jet.

To identify the maxima in the 1FGL variability flux history, we calculate the derivative at each of the 11 months in the 1FGL period. In this procedure, a peak is a point in the 1FGL curve at which the derivative changes sign.  We evaluate, for each peak, the number of points that have a positive derivative at its left plus the number of points with a negative derivative at its right. This is considered as the peak width. The sum of the absolute values of the derivatives for all the points inside the peak width constitutes the peak weight.  The larger the weight, the more significant is the peak. To avoid confusion with neighboring peaks not adequately sampled during the 1FGL period, or with flickering, we focus our analysis on the peak with the largest weight (the most significant one) inside the 11 months of the 1FGL period. Hereafter, we refer to the peak with the largest weight as the \emph{$\gamma$-ray peak}.

Figures 2 to 5 show  the 37~GHz total flux density  measurements (filled circles) and the 1FGL $\gamma$-ray monthly flux curve (open triangles)  for  some of the  sources  included in this study. The individual exponential flares and the total model fits to the flux density curve are shown in Figures 2 to 6 as thin and thick lines, respectively. For each source, we associate the  $\gamma$-ray peak with the brightest ongoing radio flare, irrespective of its phase.  Such a tie  is highlighted  by a  vertical dashed line connecting  the $\gamma$-ray peak (filled triangle) with the brightest  individual radio flare  (open circle). 

Figure 2 shows the source 0235+164. The most significant flare, which also corresponds to the highest observed monthly $\gamma$-ray flux level, occurs just before the peak of the strongest 37 GHz flare. The monthly binned $\gamma$-ray light curve for 3C~345 (Fig. 3) shows  two well-defined strong peaks. On the basis of the $\gamma$-ray peak  identification method described above, the slightly weaker peak in 2008.7 is more significant than the one considered in the statistical analysis. Nevertheless,  it can be seen that both $\gamma$-ray peaks coincide  with the rising states of individual radio flares, as in the cases of both  0420-014 (Fig. 4) and 3C 454.3 (Fig. 5). However, around 2009.4 we see a well-defined $\gamma$-ray flare in 3C 454.3,  much weaker in intensity, which is not associated with any strong radio flare. This indicates that weaker $\gamma-$ray flares might be produced by  mechanisms other than disturbances propagating downstream of the jet, a possibility already suggested in \citet{anne_2003}.

Figures 2 to 5  suggest  that  strong $\gamma$-ray  events often  occur when a mm radio flare is ongoing, more specifically during the early stages of a mm flare after it has started and is either rising or peaking. However, lacking additional information, we cannot prove a causal connection (i.e., cospatiality) between, say, the strongest gamma and radio flares in 0235+164. What we \emph{can} do is to look for statistical connections in our sample: do the most significant $\gamma$-ray peaks tend to occur during a specific phase of the radio flares? To quantify the possible  connection, the phase of each mm flare has been  characterized in the same fashion as in previous works \citep{valtaoja_1995, valtaoja_1996, anne_2003}, where the phase of a rising  flare is defined as

\begin{equation}
\phi_{rising} =S^{\gamma peak}_{mm}/S_{mm}^{max}
\end{equation}
and  the phase of a decaying flare as

\begin{equation}
\phi_{decaying} =2- S^{\gamma peak}_{mm}/S_{mm}^{max},
\end{equation}
where  $S^{\gamma peak}_{mm}$ is the flux density of the radio flare at the time of the $\gamma$-ray peak  and $S_{mm}^{max}$ is the peak flux density of the radio flare.  Equations 1 and 2 correspond to the following characterization of the flares  : \emph{0} marks the beginning of the flare,  \emph{1} the peak  of the flare, and  \emph{2} the end of the flare. Thus, a mm flare is growing if its phase ranges between 0 and 1, whereas  for any flare phase between 1 and 2 the  mm flare is decaying.  

We  determined the flare phase corresponding to the most significant $\gamma$-ray peak in 45 sources (33 quasars, 11 BL~Lacs, and 1 radio galaxy). The values are given in column 4 of Table 1. In Figure \ref{fig:phase}, we compare the phase of the mm flare at the time of the $\gamma$-ray peak and the flux density of the $\gamma$-ray peak. The top panel  gives the distribution of flare phase versus the $\gamma$-ray flux density of the most significant peak during the 1FGL period. At first glance, there appears to be a  correlation  between the flare-phase and the flux density of the strongest $\gamma$-ray events,  suggesting that the  strongest $\gamma$-ray peaks tend to occur at the earliest stages of a mm flare. However, the significance of the correlation (P = 80\%) is not high enough to  be considered as a correlation but  only as a trend. Additional \fermilat\ data will be needed to prove the possible existence of a correlation.

The bottom panel of Figure \ref{fig:phase} shows the distribution of the flare phases as a box plot, where the thick line represents the median of the distribution and the box contains 50$\%$ of the sources, delimited by the upper and lower quartiles, whereas the vertical lines indicate the extremes of the flare phases.  The median of the radio-$\gamma$-ray events is at the radio phase 0.9, and the majority of the events occur during flare phases  between 0.6 and 1.25, indicating that the $\gamma$-ray flares tend to occur during the rising or peaking state of mm flares. This relationship between the flare phase and the $\gamma$-ray maxima is consistent with previous conclusions found from the \emph{EGRET} data \citep{valtaoja_1995,valtaoja_1996, anne_2003}.

\subsection{Statistical simulations}

To  assess the statistical significance of the observed flare-phase distribution, we  performed the following test. For each source, we  performed two random shifts of the  monthly binned $\gamma$-ray light curve along the time axis and then  measured the corresponding radio phases of the most significant peaks in these shifted $\gamma$-ray light curves, thus creating a random flare-phase distribution. To test  whether  our observed flare-phase distribution (bottom panel in Figure \ref{fig:phase}) reveals an intrinsic connection or  the  distribution is drawn by chance,  we start  with the assumption that the observed and the random flare-phase distributions are drawn from the same parent distribution  and test whether this assumption is valid.

On the basis of  the chance assumption, the medians of both distributions should be nearly equal. We mix up all the  flare phases from  the observed and the random distributions. We  then  select \emph{m}  of them, which we call "observed" while the remaining \emph{n} are designated as  "random". These new distributions should be almost the same as the original distributions and their median difference should be close to zero. By repeating the procedure 1000 times, we compile a distribution of the differences between the "observed" and the "random" distributions. The original median difference between the observed flare-phase and the random flare-phase  distribution is -0.93. By using the distribution built on 1000 permutations, we compute the p-value for the probability of obtaining a median difference larger (in terms of  absolute value)  than the  observed one.  The p-value computed is 0.005, which leads us to conclude that  the  observed flare-phase  distribution (shown in the bottom-panel of figure \ref{fig:phase}) is not  drawn by chance   at  a significance level of   $ 99.5\%$ .  To help implement these statistical simulations, we  used the  permutation  routines of the statistical R language \footnote{http://www.r-project.org/} and our  IDL routines.

 \begin{figure}
 \centering
 \includegraphics[width=\columnwidth]{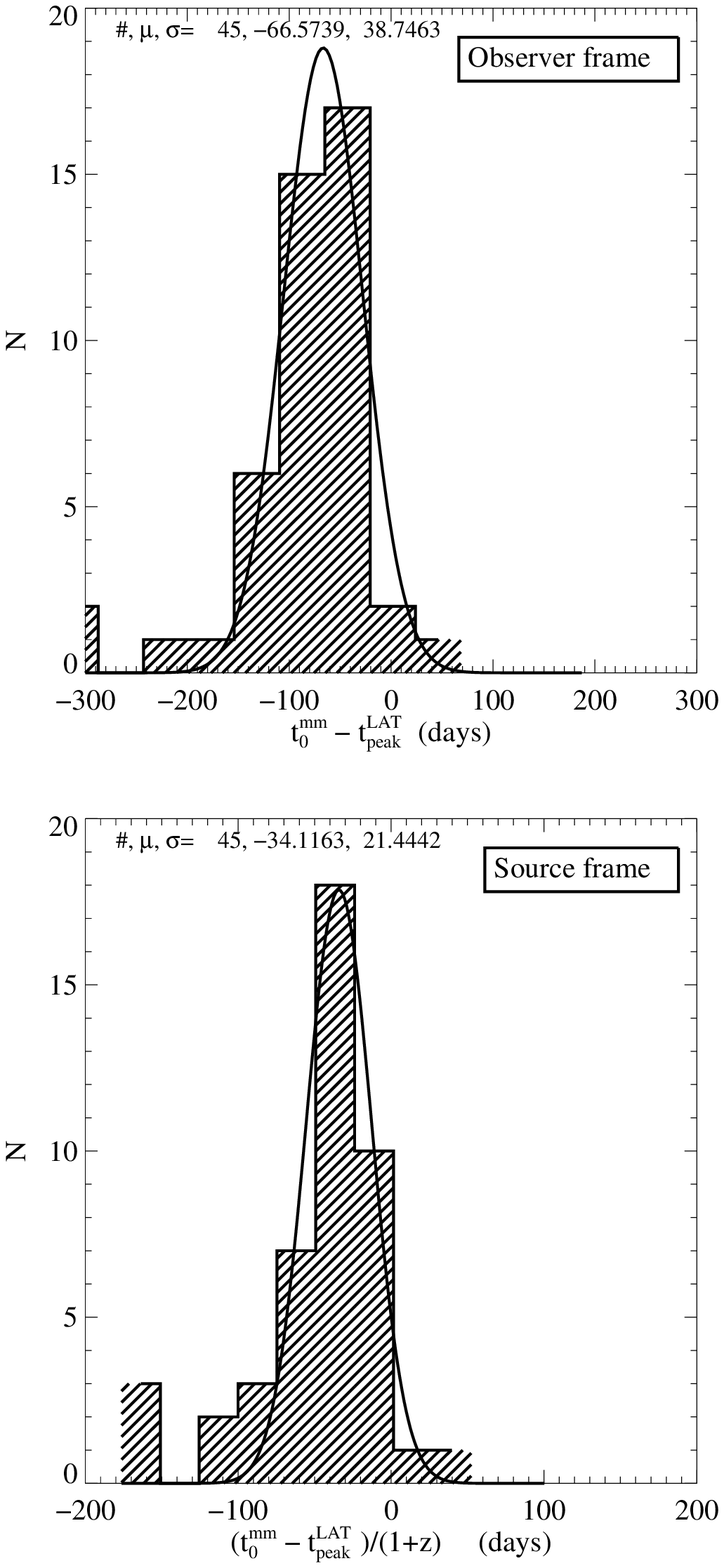}
 \caption{\emph{Top panel:} The distribution of the observed delays between the beginning of a mm flare ($t_{0}^{mm}$)  and the peak in the 1FGL light curves ($t_{peak}^{LAT}$). The mean observed delay is about 70 days, with radio leading the $\gamma$-rays. The delays for each individual source have also been estimated in the source frame and the distribution is  shown in the \emph{bottom panel}.  The mean value of the delay in the source frame is roughly one month.    }\label{fig:delay}
  \end{figure}

\section{The location of the $\gamma$-ray emission site}

The results described in the previous section   show that mm  flux densities and gamma ray  fluxes  in sources  with optical polarized emission (HPQ, LPQ) are correlated on monthly timescales, whereas for both quasars without polarization data (QSO) and  BL Lacs (BLO) such a correlation is not found. Although the inverse Compton mechanism predicts a correlation between radio and $\gamma$-ray emission strengths, studies from the \emph{EGRET} era did not find evidence of this correlation \citep[e.g.][]{anne_2003}. This apparent contradiction with results obtained here is probably due to the limited \emph{EGRET} sensitivity and to the sparse $\gamma$-ray data available at that time. The absence of $\gamma$-ray and radio flux density correlation for BL~Lacs might be an indication that $\gamma$-ray emission mechanisms or locations are different for quasars and BL~Lacs  \citep{anne_2003}. In quasars, $\gamma$-rays and radio flux densities are most significantly correlated for sources with the relativistic jet pointing close to our line of sight (HPQs), which  implies that there is a strong coupling between the radio and the $\gamma$-ray emission mechanisms. That the correlation is seen in monthly flux density levels may imply that they have  a co-spatial origin.

Perhaps the most remarkable result is that the most significant $\gamma$-ray flux peaks occur when a mm-flare is either rising or peaking. As discussed in Section 1, this indicates a sequence of events where first a disturbance (i.e. shock) emerges from the radio core, becoming visible as an increase in the mm radio flux (and as a new VLBI component, \citep{jorstad_2001}), and \emph{after} that the $\gamma$-ray flux rises and peaks. In other words, strong $\gamma$-ray flares are produced in the same disturbances that produce the mm flares, downstream of the radio core in the relativistic radio jet. Using our data, we can estimate the time delay between the time when the mm flare starts and when the $\gamma$-rays peak for each source. We define the beginning of a mm flare to be:
\begin{equation}
t_{0}^{mm} = t_{max}^{mm} - \tau,
\end{equation}
where  $t_{max}^{mm}$ is the time of the mm flare peak and $\tau$ is the variability timescale  \citep{savolainen_2002,anne_2003}. In other words, we define the beginning of the flare as the epoch when its flux is 1/e of the maximum flux.

In the top panel of Figure \ref{fig:delay}, we plot the distribution of the observed time delays. As  can be seen, the  distribution is centered around 70 days with the beginnings of the radio flares preceding the $\gamma$-ray peaks, which agrees with  results  obtained during the EGRET era. \citet{anne_2003} found time lags between 30-70 days from the onset of a millimeter flare to the $\gamma$-ray flare, and \citet{jorstad_2001} found a mean time lag from the VLBI ejection time to the $\gamma$-ray flare to be 52$\pm$76 days.  In the bottom panel of Figure \ref{fig:delay},  we show the distribution of the delays in the source frame, with a  median delay of 30 days.

To locate the maximum $\gamma$-ray production region, we  can convert the time delays to linear distances from the region where the mm outburst begins (i.e.,  the radio core, as we  argue below)    to the region of the $\gamma$-ray production   by using the  expression \citep[e.g.][]{anne_2003,pushkarev_2010}

\begin{equation}
\Delta r = \frac{\beta_{app} c (t_{0}^{mm} - t_{peak}^{LAT})} {sin \theta ~(1+z)},
\end{equation}
where  $\theta$ and $\beta_{app}$ are the jet viewing angle and the apparent jet speed, respectively. By taking the latter values from the MOJAVE  website \footnote{http://www.physics.purdue.edu/MOJAVE/index.html} and using the  delay  computed in this work,  we were able to compute the  linear distance from  the radio core  to the region of the $\gamma$-ray production for 30 sources in our sample. These values  are given in Table 1.

If we consider only those sources for which the $\gamma$-ray peak occurs when a mm flare is  rising or peaking  ($ 0 \leq phase \leq 1.25$), then an outlier-resistant determination of the mean leads us to conclude that in our sample  the  average location of  the $\gamma$-ray emission  site   is $\langle R_{\gamma}\rangle =  7.4\pm 1.3 \ pc $ downstream from the radio core, which places  the $\gamma$-ray emission site well outside the canonical BLR ($\leq 1pc$), even without taking into account  that the radio core itself is at a considerable distance from the black hole.  A contemporaneous estimate of the location of the  $\gamma$-ray emission site  for   3C~ 279  \citep[about $10^{5}$  gravitational radii,][]{abdo_3c279_2010}  is in good agreement with the distance reported in Table 1 (13.4 pc $\sim 2 \times 10^{5}$~ R$_{S}$, assuming a black hole mass of 6 $\times 10^{8}$ M$_{\sun}$). Furthermore,  \citet{agudo_2011} found that the $\gamma$-ray emission in OJ~287 is  generated  in a stationary feature located at a distance $> 14~pc$ from the central engine, which  is also consistent with the distance derived in this work ($\sim 12~pc$).  For 3C~345, \citet{schnitzel_2010} find that $\gamma$-rays are produced up to 40 pc from the engine, which is again consistent with our finding of an emission site downstream of a radio core, itself at a considerable distance from the black hole.

We note that we have compared the time delay at the beginning of the radio flare with that at the peak of the $\gamma$-ray flare, whereas ideally we should compare the beginnings of both. At present, these \emph{Fermi} $\gamma$-ray data are unavailable. However, it is well known from both  EGRET observations and the  \emph{Fermi} data analyzed here that the strongest $\gamma$-ray flares tend to  have a short duration, with typical timescales of a month or less (see also Figs. 3 and 4). The delays from the onsets of the mm-flares to the onsets of the $\gamma$-ray flares are therefore somewhat shorter than those shown in Fig. 7, but the temporal order of events is unchanged.

A number of other studies have also found that the radio core is located parsecs or even tens of parsecs away from the central engine. A partial list includes \citet{jorstad_2007}, \citet{ chatterjee_2008}, \citet{marscher_2008},  \citet{sikora_2008}, \citet{jorstad_2010}, and \citet{marscher_2010}. Thus, any correlation between $\gamma$-ray flares and radio events (millimeter flares, VLBI ejections) implies a "distant" origin scenario, with $\gamma$-rays produced far beyond the BLR, in the vicinity of the radio core situated around   $10^{5}$~R$_{S}$, or downstream from it.

However, before we can claim that all  $\gamma$-ray flares are generated at distances of parsecs downstream of the radio core, we recall the following two caveats:  
\begin{enumerate}
 \item Our results refer only to  the most prominent $\gamma$-ray peak  in the 1FGL light curves. For many sources in our sample, the 1FGL light curve displays several other flares, such as the one around the middle of 2009 in Figure 5. It is striking  that this flare occurred  during a period of quiescent mm-activity. It is possible that these smaller flares in quasars, and flares in general in BL Lacs, are produced by a different mechanism or at different locations, most likely upstream of the radio core. \\
  \item  Monthly bins in the 1FGL curves might mask  rapid $\gamma$-ray flares, or blend them with each other.  However,  even if some of our "most significant" flares result from a superimposition of two or more rapid flares, the time difference between them must be smaller than the delays found  in Table 1, and our basic argument about the connection to the radio core remains valid. The analysis of rapid  $\gamma$-ray flares is beyond the scope of this paper, but we note that significant efforts in this direction have already  been made  \citep{marscher_2010,tavecchio_2010}.  
 \end{enumerate}

Despite the above caveats, no argument conflicts with  our main conclusion, namely,  that strongest $\gamma$-ray flares tend to occur at initial or peaking stages of a mm flare and  that they are therefore  related to disturbances (i.e. shocks) propagating downstream from the core. However, both the radio core and the disturbances have finite lengths, which  complicates the interpretation. There is no commonly accepted physical model for either the radio core or  the shocked jet flow, but we can consider the following general scenario, in which a finite-length disturbance passes through an extended radio core.  We refer to Figure 9 of \citet{marscher_2009} for a pictorial representation of the AGN structure assumed here.  There are three stages during which $\gamma$-rays may be produced.

\begin{enumerate}[label=(\roman{*})]
\item \emph{Upstream of the radio core:} Once a disturbance is produced close to the black hole and the accretion disk, it
traverses downstream along the so-called  acceleration and collimation zone (ACZ)  \citep{marscher_2008}. The accretion disk and the BLR are copious sources of seed photons, which, however, diminish quickly as the disturbance approaches the radio core at the distance of several parsecs (of the order of  $10^{5}$~R$_{S}$). Eventual $\gamma$-ray flaring should clearly precede any radio events. For example, for 3C 279 \citet{chatterjee_2008} find that X-rays, presumably generated close to the central engine, lead radio variations by 130 days. For $\gamma$-rays created from accretion disk or BLR seed photons (the "near" origin scenario), comparable delays to radio variations should be seen.    
 \\
 \item \emph{Passing through the radio core:}  After the ACZ,  the disturbance  reaches a recollimation or standing conical shock \citep[][]{daly_1988,gomez_1995}, which can be most readily identified  in VLBI maps with  the brightest  stationary  feature, the so-called  \emph{radio core}. The radio emission presumably starts to rise as the moving disturbance  enters the radio core. By definition, the VLBI ejection epoch, determined by extrapolating the component motion backwards in time, should occur when the disturbance passes the center of the radio core. The relative temporal sequence of mm-flare onset, VLBI ejection epoch, and possible $\gamma$-ray flaring depend on the as yet unknown physics of shock inception and growth. However, owing to the relatively small size of the radio core no large delays between core-related events are expected. According to   \citet{marscher_2009}, the size of the radio core  can be approximated as $ a_{core}= 0.05 (43GHz/ \nu_m)$ milliarcsec,    where $\nu_{m}$ is the frequency at which the core becomes self-absorbed.  Assuming $\nu_{m} = 43$~GHz for  a typical source in our sample ($<z> \sim 0.8$,  $\beta_{app} \sim 11.1$, $\theta \sim 3^{\circ}$),  then  the passage time of a disturbance through the core  ($a_{core} \sim 0.05$ mas) is about 70 days, which  is similar  to the mean time-delay from the onset of a mm flare to  the  $\gamma-$ray peak ($\sim$2~months).    However, it must be borne in mind  that the core size here is an upper limit estimated from model fitting the unresolved stationary component  associated with the core  on VLBI images  and  may be much larger than the actual size of the radio core. In addition, we do not know if the millimeter flare begins as the disturbance enters the radio core, or first later, as opacity and shock physics are likely to play a role.
 \\
 \item \emph{Downstream of the radio core:} As the shock emerges from the core, it  becomes detectable  in VLBI maps as a new component. According to \citet{hovatta_2008}, the median rise time for a 37 GHz flare is one year, indicating that the shock traverses a considerable distance downstream from the radio core before it enters its final adiabatic decay phase. For $\gamma$-rays originating in shocks, time delays  longer than in case (ii) are expected.  We note again that a delay of months or even longer from the $\gamma$-ray flare to the peak of the radio flare does \emph{not} imply that the origin is the radio core,  and much less in the vicinity of the central engine.
  \end{enumerate}

Our main finding, that the most intense  $\gamma$-ray flares in the 1FGL light-curves occur about 70~days  after the onset of a mm flare,  indicates a "distant" origin scenario, with emission sites located  several parsecs away from the central engine. In this scenario, corresponding to cases (ii) and (iii), the highest levels of $\gamma$-ray emission are achieved  during, or after,  the passage of the disturbance through the core (or through an outer stationary  feature)\footnote{Stationary features  located downstream  from the core may play a major role in the production and release of energy in parsec-scale jets \citep[see][]{arshakian_3c3903, leontavares_3c120, lobanov_2010,agudo_2011} },  thus compressing the moving material  and enhancing substantially the energy of electrons,  leading to the  inverse Compton scattering of  low-energy photons provided  either  by the jet itself or from external photon fields.

The two-month average delay we find is most consistent  with cases (ii) and (iii), however,   for several reasons we cannot decide on whether the  strongest $\gamma$-ray flares occur  during disturbances passing through the radio core (case (ii)) or if they  originate  from the growing shocks  downstream of the radio core  (case (iii)).  First, our time resolution of one month is insufficient for detailed studies of the sequence of events during the passage of the disturbance through the radio core. Second, as can be seen in Figure 7, the spread in the lengths of the delays is not inconsiderable. Third, we have defined the beginning of the radio flare using a simple formula (Equation 3), which obviously is an approximation. Inspection of individual mm-flares in the Mets\"ahovi monitoring program reveals that in several cases the radio flux starts to increase somewhat  before or after the epoch given by Equation 3. Multifrequency monitoring, preferably combined with VLBI and  time-resolved $\gamma$-ray light curves,  is needed to achieve more accurate time determinations and studies of case (ii). 

For the strongest $\gamma$-ray flares, case (i) seems to be excluded by our results, although our time resolution cannot exclude the possibility that  some of the flares are produced a short distance (as compared to the distance to the central engine) upstream of the radio core. However, we should also remain open to the possibility that $\gamma$-ray flares are generated close to the black hole  due to an enhancement  of the local seed photons  (either from  the accretion-disk corona system or the canonical BLR), as has been  suggested  in other studies  \citep[e.g.][]{marscher_2010,tavecchio_2010}.  Nevertheless,  $\gamma$-ray flares produced in the  close dissipation  scenario ($\leq 1$ pc)  should, in general, occur months before the mm-flare inception and the VLBI  component ejection.  Whether close-dissipation $\gamma$-ray  flares are related to  weak and/or rapid  flares  will be  addressed  in a future work by means of  high-resolution $\gamma$-ray light curves.

In conclusion, the results presented in this paper strongly indicate that at least for the strongest $\gamma$-rays the production sites are downstream or within the  radio core, well outside the BLR at distances of several parsecs or even tens of parsecs from the black hole and the accretion disk. A number of papers based on  \emph{Fermi} data have reached similar conclusions, mainly for individual sources \citep[e.g.][]{abdo_3c279_2010,schnitzel_2010,agudo_2011}. In particular, Pushkarev et al. (2010) have also studied a large sample of sources statistically, concluding that the $\gamma$-ray production takes place within the extended radio core. The finding that $\gamma$-rays lead the radio emission by 1.2 months (in the source frame), which is consistent with our results when we take into account the average time delays between 37 GHz and 15 GHz and  that they compared the peak times, not the beginnings of the radio flare as we have done (cf. Section 1).

In the current AGN paradigm, the  sources of seed photons   to produce $\gamma$-rays   at distances  well beyond the canonical BLR  could be provided  by  (a) the  dusty torus \citep[e.g.][]{blazejowski_2000} or  (b)  by the jet itself, either via a  slow sheath  surrounding the fast spine of the jet \citep{ghisellini_2005},  or from the same disturbances that produce the radio outburst.  Firm detections of  IR emission associated with a  hot dusty torus have so far been limited to a couple of bright  blazars \citep{Turler_2006, Malmrose_2011}, and, as is well known, SSC models in general fail to produce the observed amounts of $\gamma$-ray emission, assuming physically reasonable parameter values for the shocked regions. However,  \citet{leontavares_3c120} and \citet{arshakian_3c3903}  have shown  that radio-loud sources with  superluminal motions  may power an additional component of  the BLR, which can be located parsecs downstream of the canonical BLR ($< 1~pc$). In effect, the flow drags a part of the BLR with it.  Based on these  results,  a tentative  idea to test  is  \emph{ whether an  outflowing BLR  can serve  as a source of external photons to produce $\gamma$-rays, even at distances of parsecs down the jet} . In this scenario, the strong $\gamma$-ray events are produced in the  shocks of the jet by upscattering  external photons  provided by the outflowing BLR.  Although  a combination of optical spectral-line monitoring  with regular VLBA observations has strongly suggested that an outflowing parsec-scale BLR  is present \citep{leontavares_3c120, arshakian_3c3903}, it is obviously of greatest importance  to extend these observations to many more objects to  explore the feasibility of producing $\gamma$-rays far downstream of the central engine and the radio core by means of EC radiation mechanisms.

The  most effective way to explore the above  scenario (and others) is  by modeling  simultaneous, well-sampled   spectral energy distributions (SEDs).  Although these high quality SEDs are quite expensive in terms of manpower and observing time, significant progress has already been made \citep[e.g.][]{fermi_sed_2010}. Even higher quality SEDs are becoming available from \emph{Planck,  Swift}, and  \emph{Fermi}  satellites and supporting ground-based observatories  \citep{planck_metsahovi_2011}.  Furthermore, the black hole mass is a crucial parameter  that controls the accretion rate. It is also found  that the more massive the black hole is, the faster and the more luminous jet it produces   \citep{valtaoja_2008,leontavares_bhm_2011}. Thus,  a reliable estimate of  the black hole masses  is an essential input  to theoretical models of both the shape and the variability of blazars SEDs.

\section{Conclusions}

We have compared monthly $\gamma$-ray light curves with the finely sampled  Mets\"ahovi 37~GHz light curves  of northern sources  identified in  the 1FGL, finding that:

\begin{enumerate}
 \item A significant correlation exists between simultaneous $\gamma$-ray  photon fluxes and 37~GHz flux densities in quasars, more specifically, among those with high optical polarization. The absence of any similar correlation for  BL~Lacs might  indicate that different   $\gamma$-ray emission mechanisms operate in  quasars and BL~Lacs.
 \\
 \item Statistically speaking, the strongest $\gamma$-ray flares occur during the rising /peaking stages of millimeter radio flares. This result supports the scenario where the $\gamma$-ray emission in blazars originates in the same disturbances in the relativistic  plasma (i.e. shocks) that produce the radio outbursts, around, or more likely downstream, of the radio core and far outside the classical BLR.   
\\
 \item The average observed time delay from the onset of the millimeter flare to the peak of the $\gamma$-ray flare is 70~days, corresponding to an average distance of 7 parsecs along the jet. At  these distances, well beyond the canonical BLR, the seed photons could originate either from the jet itself, from  a dusty torus, or from an outflowing BLR. The existence of a nonvirial, outflowing BLR can make EC models possible even at distances of parsecs down the jet.
   
\end{enumerate}

\begin{acknowledgements}
We thank the anonymous referee for pointing out the importance of considering the radio core (case (ii) above) in more detail.  We acknowledge the support from the Academy of Finland to our AGN  monitoring project (project numbers 212656, 210338, 122352 and others). This work  is related to  the International Team collaboration 160 sponsored by the International Space Science Institute (ISSI) in Switzerland. This research has made use of data from the MOJAVE database that is maintained by the MOJAVE team (Lister et al., 2009, AJ, 137, 3718). 
\end{acknowledgements}

\bibliographystyle{aa}
\bibliography{ms}

\end{document}